\documentclass[12pt]{article}

\usepackage{scicite}

\usepackage{times}

\usepackage{bm}

\usepackage{graphicx}
\usepackage{amsfonts}

\usepackage{amsmath, amsthm, amssymb}

\usepackage{color}

\newenvironment{sciabstract}{%
\begin{quote} \bf}
{\end{quote}}

\newcounter{lastnote}
\newenvironment{scilastnote}{%
\setcounter{lastnote}{\value{enumiv}}%
\addtocounter{lastnote}{+1}%
\begin{list}%
{\arabic{lastnote}.}
{\setlength{\leftmargin}{.22in}}
{\setlength{\labelsep}{.5em}}}
{\end{list}}

\newcommand{\gtappr}{{{\lower4pt\hbox{$>$} } \atop \widetilde{ \ \ \ }}}
\newcommand{\ltappr}{{{\lower4pt\hbox{$<$} } \atop \widetilde{ \ \ \ }}}

\newcommand{\beq}{\begin{equation}}
\newcommand{\eeq}{\end{equation}}

\newcommand{\ud}{\mathrm{d}}
\newcommand{\pd}{\partial}

\newcommand{\kB}{k_\text{B}}

\newcommand{\ybal}{$\beta$-YbAlB$_4\,$}
\newcommand{\CeCuAu}{CeCu$_{6-x}$Au$_x\,$}
\newcommand{\yrs}{YbRh$_2$Si$_2\,$}

\def\gsim{\buildrel {\textstyle >}\over {_\sim}}
\def\lsim{\buildrel {\textstyle <}\over {_\sim}}

\newsavebox{\fmbox}

\title{Quantum Criticality without Tuning in the Mixed Valence Compound \ybal}

\author
{Yosuke Matsumoto$^1$, Satoru Nakatsuji$^{1\ast}$, Kentaro Kuga$^1$,\\ Yoshitomo Karaki$^{1\dag}$,
 Naoki Horie$^1$, Yasuyuki Shimura$^1$,\\ Toshiro Sakakibara$^1$, Andriy H. Nevidomskyy$^{2,3}$, Piers Coleman$^{2,4\ast}$\\
\\
\normalsize{$^{1\,}$Institute for Solid State Physics, University of Tokyo, Kashiwa 277-8581, Japan}\\
\normalsize{$^{2\,}$Center for Materials Theory, Department of Physics and Astronomy, Rutgers University,}\\
\normalsize{Piscataway, N.J. 08854, USA}\\
\normalsize{$^{3\,}$Department of Physics and Astronomy, Rice University, Houston, Texas 77005, USA}\\
\normalsize{$^{4\,}$Department of Physics, Royal Holloway, University of London, Egham, Surrey TW20 0EX, UK}
\\
\normalsize{$^\ast$To whom correspondence should be addressed. E-mail: satoru@issp.u-tokyo.ac.jp;}\\
\normalsize{coleman@physics.rutgers.edu.}\\
\normalsize{$^{\dag\,}$Present address: Faculty of Education, University of the Ryukyus, }\\
\normalsize{Nishihara, Okinawa 903-0213, Japan}\\
}

\date{}

\begin{document} 


\baselineskip24pt

\maketitle 

\newpage


\begin{sciabstract}
Fermi liquid theory, the standard theory of metals, has been
challenged by a number of observations of anomalous metallic behavior
found in the vicinity of a quantum phase transition. The breakdown of
the Fermi liquid is accomplished by fine-tuning the material to a quantum
critical point using a control parameter such as the
magnetic field, pressure, or chemical composition.  Our high precision
magnetization measurements of the ultrapure $f$-electron based
superconductor \ybal demonstrate a scaling of its free energy
indicative of zero-field quantum criticality without tuning in a
metal.  The breakdown of Fermi-liquid behavior takes place in a
mixed-valence state, in sharp contrast with other known examples of
quantum critical $f$-electron systems that are magnetic Kondo lattice
systems with integral valence.
\end{sciabstract}


\newpage

Quantum phase transitions occur at zero temperature as a consequence
of quantum, rather than thermal correlations.  Generally, a quantum
critical point (QCP) can be reached by driving a finite temperature
critical point~\cite{Lohneysen07,gegenwart08} or a first-order
critical end-point~\cite{SrRuO-327-endpoint} to absolute zero.  The
breakdown of Fermi liquid (FL) behavior in metals observed near a
magnetic QCP challenges our current understanding of strongly
correlated electrons. While the mechanism of unconventional quantum
criticality is actively debated, there is a growing
consensus that the underlying physics involves a jump in the Fermi
surface volume associated with a partial electron localization
\cite{Schroder00,Coleman01,Si01,Senthil03,Paschen04}. 
To date, the FL breakdown has only been observed by fine-tuning a
material to a QCP using a control parameter such as magnetic field,
pressure, or chemical composition.

Recent work, reporting the discovery of superconductivity in an
ytterbium-based heavy fermion  material \ybal, 
raised the interesting possibility that this system may exhibit quantum
criticality without tuning\cite{nakatsuji08}. 
In this compound, 
signatures of quantum criticality were observed to develop above a tiny
superconducting (SC) dome, with a SC transition temperature
of $T_{c}\sim 80$ mK  and an upper critical field
$\mu_0H_{c2}\approx 30$~mT\cite{nakatsuji08, KugaPRL}. While this observation motivated the
possibility of a zero field quantum critical point, it did not 
rule out a QCP located near
the upper critical field $H_{c2}$, as observed in the heavy fermion
superconductor CeCoIn$_{5}$~\cite{Bianchi03}.


In this report, we present clear and quantitative
evidence that quantum criticality develops at zero field
without tuning in \ybal, buried deep inside the SC dome. Moreover, we
report a simple $T/B$-scaling form of the free energy spanning 
almost four decades in magnetic field, revealing that the signatures of the
putative quantum critical point extend up to temperatures $T$ and fields $B$
more than 100 times larger than $T_{c}$ and $\mu_0 H_{c2}$, respectively.

To quantify the free energy $F(T,B)$, we employed high-precision
measurements of the magnetization $M=-\pd F/\pd B$. 
Measurements were made on ultra high purity single crystals with a
mean free path exceeding 1000~\AA\ and residual resistivity less than
$0.6~\mu\Omega$ cm, which were carefully etched to fully remove
surface impurities~\cite{MM}.
Our measurements revealed a simple $T/B$ scaling over a wide range of
temperature and field, governed by single quantum-critical (QC) 
scaling exponent previously masked~\cite{nakatsuji08} by a
limited experimental resolution and the impurity effects 
caused by surface and bulk impurities~\cite{MM}.
The $T/B$ scaling leads to the following significant consequences. 
First, the quantum critical physics is 
self-similar over four decades of $T/B$, with no intrinsic energy scale. 
Second, the field-induced Fermi liquid is characterized by a  Fermi temperature that grows
linearly with the field, determined by the Zeeman energy of the underlying critical modes.
Finally, the scaling allowed us to determine an upper bound on the magnitude of the critical field 
${\vert B_c \vert <0.2}$~mT, which is well inside the SC dome and comparable with the Earth's magnetic field: this indicates that \ybal is intrinsically quantum critical, without tuning the magnetic field, pressure, or composition.

These results are surprising, given 
the fluctuating valence nature of this material, with valence
Yb$^{+2.75}$ significantly far from integral, revealed by recent experiments~\cite{ybal-valency}.
All QC heavy-fermion intermetallics known to date have an almost
integral valence which stabilizes the local moments
\cite{Lohneysen07,gegenwart08}.  Such so-called Kondo lattice systems
are characterized by a small characteristic scale $T_0$, below which
the moments are screened to form a paramagnetic heavy FL.  Various
types of order, such as superconductivity and antiferromagnetism
(AFM), compete with the heavy FL, leading to quantum criticality, as
seen in, for example, CeCu$_{5.9}$Au$_{0.1}$ ($T_0 = 6.2$~K)
\cite{lohneysen96} and \yrs (24~K) \cite{gegenwart08}.
By contrast, mixed-valence compounds display a
much larger $T_0$, below which they typically behave as stable FLs with
moderate quasiparticle effective masses and no competing order. For example,
YbAl$_3$, with non-integral valence Yb$^{+2.71}$,
is characterized by $T_0 \sim 300$~K \cite{YbAl3}.

A remarkable feature of $\beta$-YbAlB$_4$ (Fig.~1A) is that it is quantum critical~\cite{nakatsuji08}, yet the
scale $T_0 \sim 250$~K, 
obtained from the resistivity coherence peak, is one or two orders of magnitude larger than in
other known QC 
materials. This is confirmed by the
scaling behavior of the magnetic specific heat: ${C_{\rm M}}/{T} =
\frac{S_0}{T_0}\ln\left({T_0}/{T} \right)$, where $S_0$ is a constant (inset of Fig.~1B). 
The $-\ln T$ dependence of ${C_{\rm M}}/{T}$ in the three QC materials
CeCu$_{5.9}$Au$_{0.1}$ \cite{lohneysen96}, \yrs
\cite{gegenwart08,Custers03}, and \ybal collapse onto one curve after setting
$T_0$ for \ybal $\sim 200$~K. The recent observation of intermediate
valence (Yb$^{+2.75}$) in \ybal at 20 K using hard X-ray photoemission
spectroscopy~\cite{ybal-valency} is consistent with this large 
$T_0$.

The quantum criticality in this valence fluctuating state is
accompanied by several properties reminiscent of an integral-valence Kondo
lattice. To understand their origin, it is useful to compare them
with those of $\alpha$-YbAlB$_4$ (Fig.~1A)~\cite{Macaluso07}, which is a locally isostructural polymorph of \ybal and a FL with a
similarly intermediate valence (Yb$^{+2.73}$)~\cite{ybal-valency}. 
Instead of Pauli paramagnetism normally
seen in a valence fluctuating material, the magnetic susceptibilities
of both materials display Curie-Weiss behavior with Weiss temperature $\Theta_{\rm W}$, $\chi = C/(T+\Theta_{\rm W})$, indicating the existence of local moments
(Fig.~2A). In addition, both materials have a maximum in $-dM/dT$ at $T^* \sim 8$~K, signaling a crossover from local moment behavior (Fig.~S1). Below $T^*$, $C_{\rm M}/T$ of the $\alpha$ 
phase levels off to a constant characteristic of heavy FL behavior,
whereas that of the $\beta$ phase continues to diverge (Fig. 1B).
Thus, the fate of local moments found above $T^*$ is different in these locally isostructural systems: Yb spins are fully screened in the $\alpha$ phase, but may well survive down to lower temperatures in the $\beta$ phase and produce the quantum criticality.  In both phases, strong correlation
effects are manifest, for example, in the strongly enhanced
$\left.\frac{C_{\rm M}}{T}\right|_{T\to0} \gsim 130 $~mJ/mol K$^2$,
two orders magnitude larger than the band calculation estimate ($\sim
6$~mJ/molK$^2$)~\cite{Andriy09}.

These signatures indicate that both phases are governed by two
distinct energy scales: a high-energy valence fluctuation scale 
$T_0 \sim 200$~K, and a low-energy scale $T^*\sim
8$~K, characterizing the emergence of
Kondo-lattice physics. A possible origin of this behavior is
the presence of ferromagnetic (FM) interactions between Yb moments,
manifested by the large Wilson ratio $R_{\rm W}$ between $\chi$ and $C_{\rm M}/T$,
observed in both $\alpha$- and $\beta$-phases~\cite{nakatsuji08} (minimum estimate $R_W\gtrsim 7$), and further corroborated by the observation of an ESR
signal~\cite{pagliuso-esr}, generally only seen in the presence of FM
correlations~\cite{fmESR}.  Such FM interactions are known to give
rise to Kondo resonance narrowing~\cite{kondo-narrowing} in
$d$-electron systems where Hund's coupling causes a marked reduction
in the Kondo
temperature~\cite{kondo-narrowing,daybell}. 

In the present case, the role of Hund's coupling is
played by FM intersite RKKY interactions, probably along the short
Yb-Yb bonds that form chains along the $c$-axis.  The moments of a few
$n$ neighboring Yb ions may thus become aligned, forming a fluctuating
``block'' spin $S = n J$. 
The observed valence Yb$^{+2.75}$ could then be understood in terms of 
Yb$^{3+}$$\rightleftharpoons$Yb$^{2+}$ fluctuations, as at any one time, approximately  
$1/4$ of the Yb atoms along the chains are in a singlet Yb$^{2+}$
configuration, forming the ferromagnetic blocks of
approximately $n\!\sim\! 3$ spins. The effect of these block spins is to
exponentially suppress the characteristic spin fluctuation
scale~\cite{kondo-narrowing}, resulting in localized-moment behavior.
The absence of long-range magnetic order in the $\alpha$ or $\beta$
phase points to the presence of competing magnetic interactions. Indeed, the Weiss temperature
$\Theta_{\rm W}\sim-110$~K (Fig. 2A), characteristic of an
AFM, indicates the importance of magnetic frustration.
The competing  interplay of FM interactions and valence fluctuations thus leads to Kondo-lattice-like behavior in a mixed valent material, setting the stage for quantum criticality to emerge at lower temperatures in \ybal. 

   
A prominent feature of the quantum criticality in \ybal is the divergence of the magnetic susceptibility $\chi$ 
as $T \to 0$. 
By examining  the field evolution of  magnetization $M=-\pd F/\pd B$ as a function of both $T$ and $B$~\cite{MM}, we can accurately probe the free energy $F$ near quantum criticality.
Figure~2A shows the $T$ dependence of $\chi(B)\!=\!M/B$ for different values of $B \parallel c$. Spanning four orders of magnitude in $T$ and $B$, the data show a systematic evolution from a non-Fermi liquid (NFL) metal with divergent susceptibility at zero field ($\chi\sim T^{-1/2}$) 
to a FL with finite $\chi$ in a field $g\mu_\mathrm{B}B\gtrsim k_\mathrm{B}T$.

Intriguingly, the evolution of $M/B$ found in the region $T \lesssim 3$~K and $B \lesssim 2$~T (see the inset of Fig. 2A) can be collapsed onto a single scaling function of the ratio $T/B$: 
\beq 
-\frac{\ud M}{\ud T} = B^{-1/2}\phi \left(\frac{T}{B} \right)
\label{dMdT} 
\eeq 
as shown in Fig.~2B. 
The peak of the scaling curve lies at $k_{B}T /g\mu_{\rm
B}B \sim 1$, marking a cross-over between the FL and NFL regions,
showing that 
$k_{B}T_{F}\sim g\mu_{\rm B} B$ plays the role of a field-induced
Fermi energy, 
as shown in the inset 
of Fig.~2B.
Integrating both parts of Eq.~(\ref{dMdT}), one obtains the following scaling law for the free energy~\cite{MM}:
\beq
F_\text{QC}= B^{3/2} f\left(\frac{T}{B}\right),
\label{free-energy}
\eeq
where $f$ is a scaling function of the ratio $T/B$ with the limiting
behavior: $f(x)\propto x^{3/2}$ in the NFL regime ($x \gg 1$) and
$f(x)\propto \mathrm{const} + x^2$ in the FL phase ($x\ll 1$). 
Indeed, the observed scaling of $\ud M/\ud T$ in Eq.~(1) is best fitted with $\phi(x) = \Lambda x(A+x^2)^{\frac{\alpha}{2}-2}$~\cite{MM}, resulting in a particularly simple form of the free energy:
\beq
F_\text{QC} =  -\frac{1}{(\kB \tilde{T})^{1/2}}\left((g\mu_\text{B}B)^2 + (\kB T)^2 \right)^{3/4}, \label{theor-free}
\eeq
with the best fit obtained with effective moment $g\mu_\text{B} = 1.94 \mu_B$ and the energy scale $\kB\tilde{T}\approx6.6$~eV of the order of the conduction electron bandwidth~\cite{MM}. 
This means that the free energy  depends only on the distance from the origin in the ($T$,$B$) phase diagram, similarly to the $T/B$ scaling established in the Tomonaga--Luttinger liquids in one-dimensional metals~\cite{giamarchi, sachdev}.
Equation~\ref{theor-free} implies that the effective
mass of the quasi-particles diverges as $m^*\sim B^{-1/2}$ at the QCP~\cite{MM}. 
This divergence in a 3D material, together
with the $T/B$ scaling, cannot be accounted for by the standard theory
based on spin-density-wave fluctuations
\cite{Moriya85,Millis93}. Instead, it indicates a breakdown of the FL
driven by unconventional quantum criticality.

The $T/B$ scaling suggests that the critical field $B_c$ of the quantum phase transition is actually zero. A finite $B_c$ would
require that the argument of the scaling functions $f(x)$ and
$\phi(x)$ is the ratio $x=T/|B-B_c|$, as seen for instance in \yrs\cite{Custers03}. To place a bound on $B_c$,
we substituted this form for $x$ into Eq.~\ref{dMdT}, seeking the
value of $B_c$ that would best fit the experimental data. 
The Pearson's correlation coefficient $R$ obtained
for this fit (inset of Fig.~2B), indicates that $B_{\rm c}$ is optimal at $-0.1 \pm
0.1$~mT.  
The uncertainty is only a few times larger than the Earth's
magnetic field ($\sim 0.05$~mT). More significantly, it is two orders
of magnitude smaller than $\mu_0 H_{\rm c2} = 30$ mT, and strikingly six orders of magnitude smaller than valence fluctuation scale
$T_0\sim 200$~K.
Thus \ybal provides a unique example of essentially zero-field quantum criticality.

Further evidence for zero-field quantum criticality is obtained 
from an analysis of the magnetocaloric ratio, $\Gamma_H\equiv -
\frac{1}{T}\frac{\pd S/\pd B}{\pd S/\pd T} = -\frac{\pd M/\pd
T}{C}$ (Fig.~3). Here, $C$ is the total specific heat~\cite{MM}. 
Our results show a clear divergence of $\Gamma_H/B$ as $T \to 0$ in
the NFL regime, which is a strong indicator
of quantum criticality~\cite{Zhu03}.
From the NFL regime, we can extract the critical field $\vert B_c \vert < 0.2$~mT, consistent with the estimate of $B_c$ obtained from the scaling behavior of $M$, Eq.~\ref{dMdT} (Fig. S2). 

The remarkably simple $T/B$ scaling  
in the thermodynamics 
enables us to 
characterize the QC excitations of \ybal.
In particular, the collapse of all magnetization data in terms of 
the dimensionless ratio  $r=k_\mathrm{B}T/ (g\mu_\mathrm{B}B)$ 
between the Boltzmann energy $k_{B}T$ and the
Zeeman energy $g\mu_{\mathrm B}B$ indicates an
absence of scale in the zero-field  normal state.
Furthermore, the appearance of a field-induced Fermi energy,  linear
over more than three decades in $B$, shows that the underlying 
critical modes are magnetic in character.

Using the Heisenberg energy-time uncertainty principle ($\Delta t
\Delta E \gtappr \hbar $), 
we can reinterpret the $T/B$ scaling in the time domain, 
visualizing  the field-induced FL 
as a kind of ``quantum  soda'' of 
bubbles of quantum critical 
matter of finite duration $\tau_Q=\hbar/g\mu_\mathrm{B}B$, immersed in
a FL.  
At finite temperatures, thermodynamics averages 
the physics over a thermal time scale $\tau_T =\hbar/k_\mathrm{B}T$, thus
the quantity $r = \tau_{Q} / \tau_T  \sim T/B$ 
in the scaling is the  ratio of the correlation time $\tau_{Q}$
to the thermal time-scale $\tau_{T}$.
At low temperatures, $r\ll1$ ($\tau_{T}\gg \tau_{Q}$), 
thermodynamics probes the FL exterior of the bubbles, 
but when 
$r\gg1$ and $\tau_{T}\ll\tau_{Q}$, it reflects the QC interior of the bubbles. 
This  accounts for the cross-over between FL and QC behaviors at $r\sim 1$. Moreover,
$T/B $ scaling over a wide range $r\!\sim\!10^{-1}$ to
$\sim\!10^3$ indicates that the quantum fluctuations in the
ground-state are self-similar down to
1/1000th of the correlation time $\tau_Q$. 

The observation of zero-field quantum criticality in valence
fluctuating \ybal cannot naturally be interpreted as a conventional
QCP, which would require a fortuitous combination
of structure and chemistry to fine-tune the critical field $B_{c}$ to
within $0.2$ mT of zero.
%
%
A more natural interpretation of the results is that \ybal
forms a quantum critical phase that is driven into a FL state by an
infinitesimal magnetic field. 
The $T$/$B$ scaling requires that the critical modes are Zeeman-split
by a field, and as such, various scenarios, such as critical Fermi surfaces~\cite{senthil08} or local quantum criticality with $E/T$ scaling~\cite{Si01} may be possible contenders for the explanation, provided they can be stabilized as a phase.
Established theoretical examples of a critical phase with $T/B$ scaling include the
Tomonaga--Luttinger liquid in 
half-integer spin chains and the one-dimensional Heisenberg ferromagnet~\cite{giamarchi, sachdev} .
Experimentally, the $d$-electron metal MnSi
is a candidate for a quantum critical phase with anomalous
transport exponents observed over a range of applied
pressure~\cite{MnSi}. While present work provides a strong indication for existence of such a phase in \ybal, future studies, in particular under pressure, are necessary in order to establish it definitively. 

\bibliography{ybscal-science}

\begin{thebibliography}{10}
\bibitem{MM}
Materials and methods, as well as the details of the scaling analysis and the
  difference from the previously reported \cite{nakatsuji08} $\chi(T)$ and
  $C_{\rm M}(T)$, are available in Supporting Online Material.
\end{thebibliography}

\bibliographystyle{Science}


\begin{scilastnote}
\item We thank H. Ishimoto, D.E. MacLaughlin, K. Miyake, T. Senthil, 
Q. Si, T. Tomita, K. Ueda, and S. Watanabe for useful discussions.
This work is partially supported by Grants-in-Aid (No.
21684019) from JSPS, by
Grants-in-Aids for Scientific Research on Priority Areas (No. 19052003) and
on Innovative Areas (No. 20102007, No. 21102507) from MEXT, Japan, by Global
COE Program ``the Physical Sciences Frontier", MEXT, Japan, by Toray Science
and Technology Grant, and by a grant from the
National Science Foundation DMR-NSF-0907179 (P. C. and A. H. N.). P. C. and
A. H. N. acknowledge the hospitality of the Aspen Physics Center.


\end{scilastnote}

\noindent {\bf Supporting Online Material}\\
\noindent www.sciencemag.org\\
\noindent Materials and Methods\\
\noindent SOM Text\\
\noindent Figs. S1, S2, S3 and S4\\


\clearpage

\begin{figure}[t]
\begin{center}
\includegraphics[width=10 cm, clip]{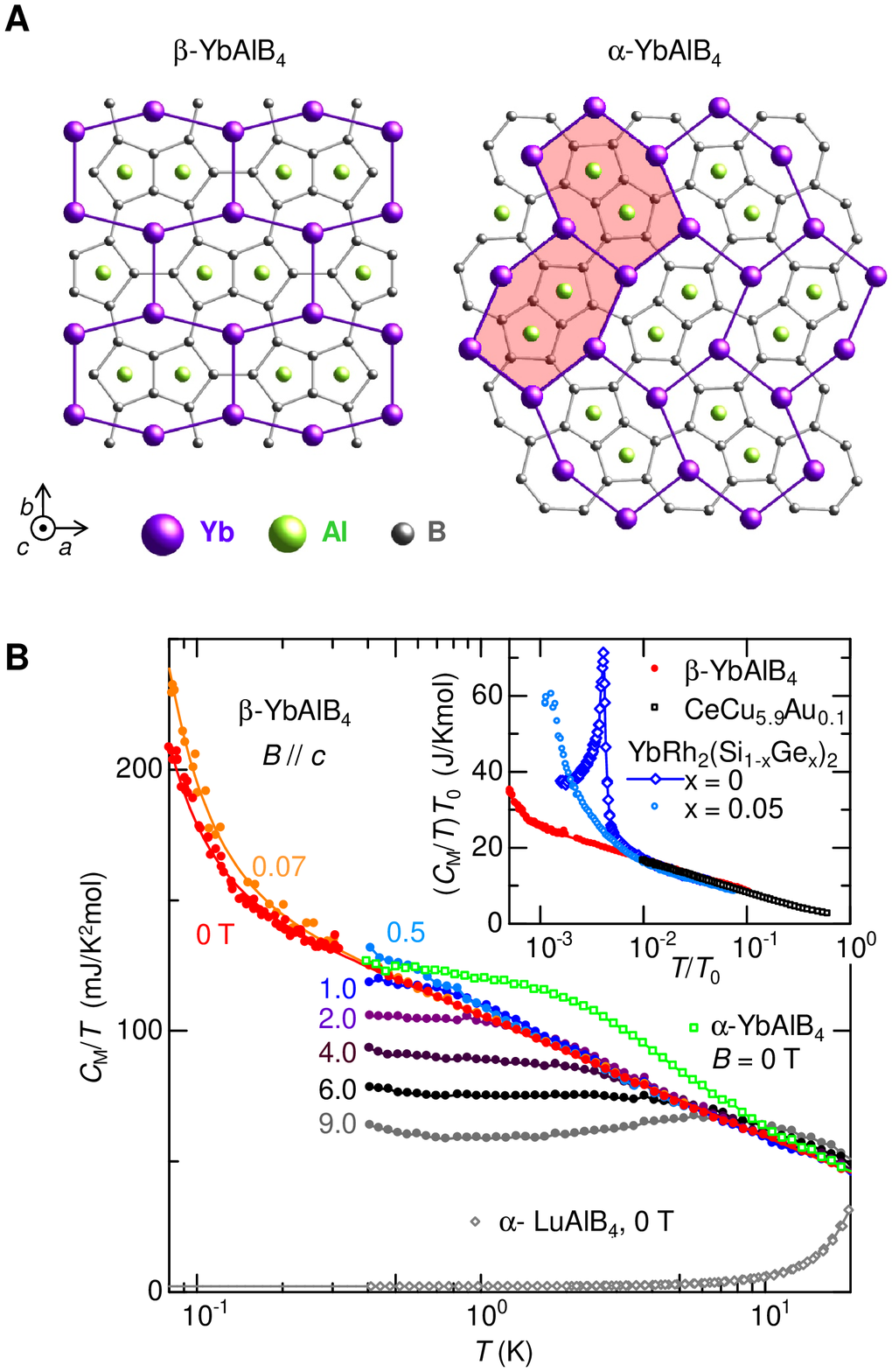}
\caption{\label{Fig1}(A) Crystal structures of $\beta$- and $\alpha$-YbAlB$_4$, which are formed from straight and zigzag arrangements of distorted hexagons of Yb atoms (shaded in red for the $\alpha$ phase), respectively \cite{Macaluso07}. The crystallographic unit cells of both phases are orthorhombic and can be viewed as an interleaving of planar B-nets and Yb/Al-layers. (B) Magnetic part ($f$-electron contribution) of the specific heat $C_{\rm M}$ plotted as $C_{\rm M}/T$ versus $T$ for both $\beta$- (solid circles) and $\alpha$-YbAlB$_4$ (open squares)~\cite{MM}. $C_{\rm M}/T$ at $B = 0$ for the $\beta$ phase shows a $\ln T$ dependence 
for 0.2 K $< T <$ 20 K. $T_0 \sim$ 200 K was determined from the fit to ${C_{\rm M}}/{T} = {S_0}/{T_0}\ln\left({T_0}/{T} \right)$. 
The upturn in the lowest $T$ may contain a nuclear contribution. 
Inset
: $C_{\rm M}/T$ scaled by $T_0$ compared with quantum critical systems 
CeCu$_{5.9}$Au$_{0.1}$ ($T_0$ = 6.2 K)\cite{lohneysen96} and YbRh$_2$(Si$_{1-x}$Ge$_x$)$_2$ ($T_0$ = 24 K) \cite{gegenwart08,Custers03}. The $\ln T$ dependence of the three quantum critical materials collapse on top of each other using nearly the same
coefficient $S_0 \sim 4$ J/mole K, indicating a common meaning of $T_0$ as the $T$ scale below which $\sim$ 70~\% of the ground doublet entropy, $R \ln2$, is released.}
\end{center}
\end{figure}

\begin{figure}[t]
\begin{center}
\includegraphics[width=9 cm, clip]{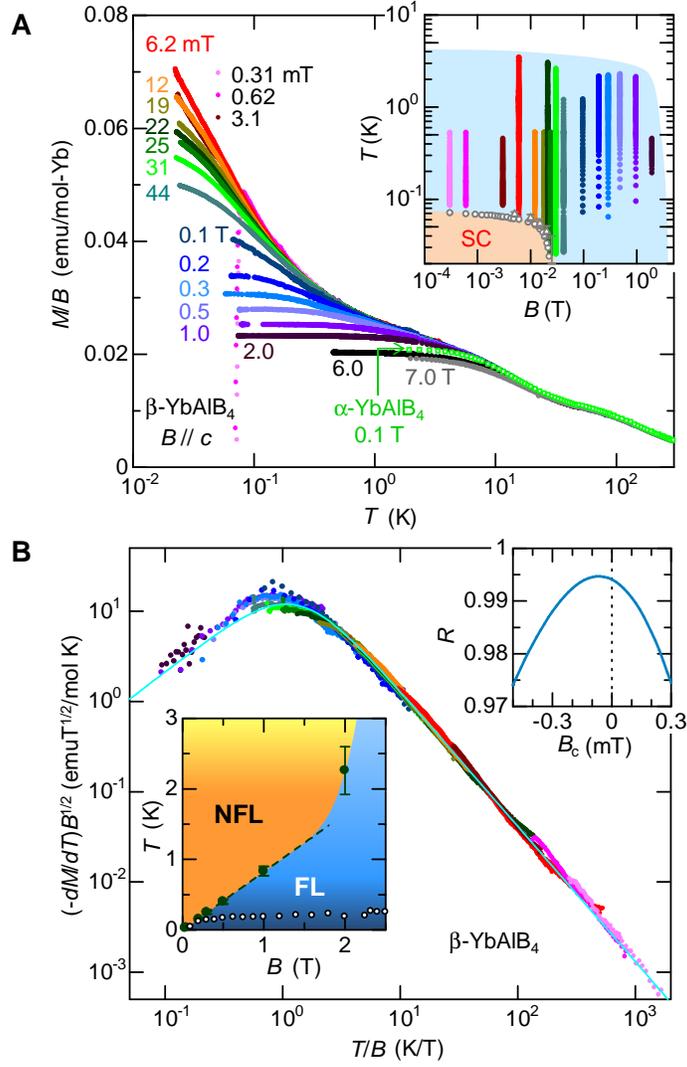}
\caption{\label{Fig2}(A) Temperature dependence of the magnetic susceptibility 
$M/B$ of both $\beta$- (solid circles) and $\alpha$-YbAlB$_4$ (open squares). The Curie-Weiss fit above 150 K yields a Weiss temperature $\Theta_{\rm W} \sim -110$ K and an effective moment of $\sim 2.2 \mu _B$ for both systems. 
Inset shows the quantum critical $B$-$T$ range where the scaling applies (solid circles in the blue shaded region) and the superconducting (SC) phase under the upper critical fields (open circles and triangles, see \cite{MM}). (B) Scaling observed for the magnetization at $T \lsim 3$ K and $B \lsim 2$ T. The data was fitted to the empirical Eq. 1 with scaling function $\phi(x) = \Lambda x (A+x^2)^{-n}$, a form chosen to satisfy the appropriate limiting behavior 
 in the Fermi liquid regime~\cite{MM}. The right inset shows Pearson's correlation coefficient $R$ for the fit with finite $B_c$. Note that $R$ reaches a maximum value of 1 if the fit quality is perfect. The best fit is obtained with $n = 1.25 \pm 0.01$ and $B_c = -0.1 \pm 0.1$ mT (light blue line), corresponding to $\alpha=3/2$ in the scaling form of the free energy, Eq.~2, see \cite{MM}, and $\vert B_c \vert< 0.2$ mT. The left inset shows the $B$-$T$ phase diagram of \ybal in the low $T$ and $B$ region. The filled circles are determined from the peak temperatures of $ -dM/dT$ below which the FL ground state is stabilized. At low field, the thermodynamic boundary between the FL and NFL regions is on a $k_{B}T \sim g\mu_{\rm B}B$ line (broken line). 
The open circles are the temperature scale $T_{\rm FL}$ below which the $T^2$ dependence of the resistivity is observed\cite{nakatsuji08}.}
\end{center}
\end{figure}

\begin{figure}[t]
\begin{center}
\includegraphics[width=8 cm, clip]{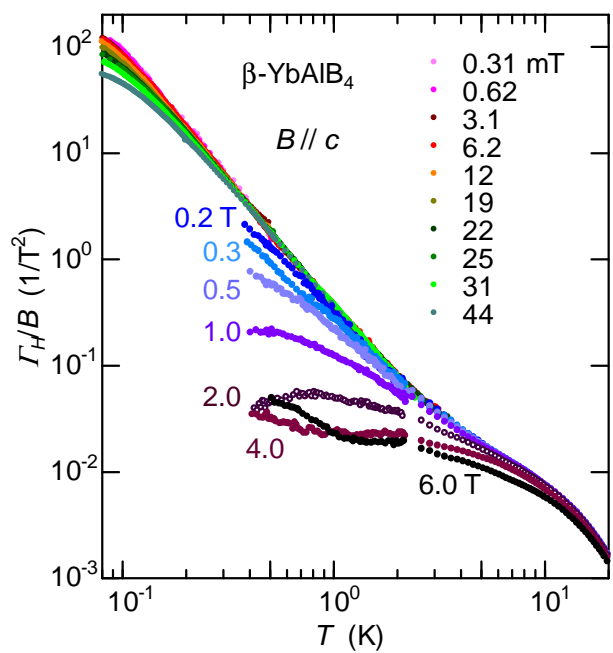}
\caption{\label{Fig3}Temperature dependence of the magnetocaloric effect divided by $B$, $\Gamma_H/B$. See also Fig.~S2 for field dependence of $\Gamma_H$.}
\end{center}
\end{figure}

\clearpage

\renewcommand{\figurename}{Figure S\!\!}
\setcounter{figure}{0}

\section*{Supporting Online Material}
\subsection*{Materials and methods}

High purity single crystals of \ybal were grown by a flux method \cite{beta1}. Energy dispersive x-ray analysis found no impurity phases, no inhomogeneities and a ratio Yb:Al of 1:1. Surface impurities were carefully removed with dilute nitric acid before the measurements. A large mean free path exceeding 1000 \AA\ obtained from Shubnikov-de Haas measurements strongly supports the high purity of the samples~\cite{Eoin_PRL09-S}.

The magnetization data at $T < 4$ K and $B < 0.05$ T were obtained by using a high precision SQUID magnetometer installed in a $^3$He-$^4$He dilution refrigerator \cite{matsumoto}. For 
fields above 0.05 T, we utilized a high precision Faraday magnetometer which is also installed in a $^3$He-$^4$He dilution refrigerator \cite{sakakibara}. These magnetometers have a resolution of $\sim 10^{-8}$ emu for the SQUID and $\sim 10^{-5}$ emu for the Faraday system. The samples used for the SQUID magnetometer measurements were high-purity single crystals with the residual resistivity ratio (RRR) $>$ 200 ($\sim$ 30 pieces, 0.82 mg). These samples were mounted using silver paste and were inserted in a pick-up coil and cooled by a heat link made of silver foils. 
The superconducting magnet was covered with a Nb superconducting shield and a $\mu$-metal tube in order to eliminate the Earth's magnetic field. The pick-up and primary lines covered with Pb superconducting tubes were connected to a multiple purpose dc-SQUID probe located in a bath of liquid $^4$He at 4.2 K. The magnetization $M$ and ac-susceptibility $\chi _{ac}$ were obtained as dc and ac signals of the SQUID output. $\chi _{ac}$ below 1 mT did not show any field dependence in the temperature range $T > 0.08$ K. This allows us to use the $\chi _{ac}$ data obtained at 0.31 mT and 0.62 mT to estimate the magnetization at the corresponding fields in the scaling analysis. For the $\chi _{ac}$ measurement, an ac field of 0.1 $\mu$T and 
frequency $f=$ 16 Hz is applied along the $c$-axis. The residual magnetic field was estimated to be $\sim$ 1.1 $\mu$T using the dc-diamagnetic signals of the superconductivity of \ybal measured under various magnetic fields of the order of $\mu$T. For the Faraday magnetometer measurements, we used single crystals of 7.5 mg, 
whose typical RRR is as high as 140. The absolute values of $M$ for both measurements were calibrated by comparing the data with those measured by a commercial SQUID magnetometer at $T \ge 2.0$ K. 

The specific heat $C$ was measured by a relaxation method. For the measurements above 0.4 K, high-purity single crystals of 0.8 mg with RRR $>$ 200 were measured by using a physical property measurement system (PPMS). The data below 0.4 K were obtained for high-purity single crystals of 2.2 mg with RRR $>$ 200 by using a heat capacity cell installed in a $^3$He-$^4$He dilution refrigerator. The magnetic part of the specific heat $C_{\rm M}$ was obtained by subtracting the specific heat of $\alpha$-LuAlB$_4$, which is the non-magnetic isostructural counterpart of $\alpha$-YbAlB$_4$ \cite{beta1}. To estimate $C$ at $B < 0.07$ T in the analysis of $\Gamma_H$, a linear interpolation of the specific heat data at $B = 0$ and 0.07 T was used (Fig.~1B in the main text). The improved crystal quality, measurement technique, and more careful procedure to remove surface impurities of the crystals allowed us to measure more precisely the 
values of the low-temperature specific heat coefficient (e.g. $C/T \sim 130$~mJ/mol$\cdot$K$^2$ at $T=0.4$~K), which are fully reproducible and should replace the somewhat inaccurate values reported earlier in Ref.~S5. 
Actually, in thermodynamic measurements in Ref.~S5, 
a batch of about 50 crystals had to be used to gain signal intensity, 
with typical RRR values between $\sim$10 and 300.

\subsection*{Supporting online text}
\subsubsection*{1. Temperature dependence of the magnetizaiton and magnetocaloric ratio}

In order to characterize the power law behavior in the temperature dependence of the magnetization, we show
 $-dM/dT$ versus $T$ for selected fields along the $c$-axis on a logarithmic scale in Figure S1 for both $\beta$- (solid circle) and $\alpha$-YbAlB$_4$ (open square). Below $T\sim$ 3 K, $-dM/dT$ for the $\beta$ phase shows $T^{-1.5}$ dependence, indicating the divergence of the susceptibility $\chi\sim T^{-1/2}$ (broken line) in the non-Fermi liquid region at $T \gtrsim B$, and $T$-linear behavior (solid line) as expected for a Fermi liquid at $T \lesssim B$.
The derivative of the magnetization $-dM/dT$ under $B = 6$ T for the $\beta$-phase and under $B = 0.1$ T for the $\alpha$-phase both show the peak at $T^* \sim 8$ K, due to the crossover from the high temperature local moment behavior to the Fermi liquid behavior at low temperatures. 

The magnetocaloric ratio is defined as
\beq\label{S.GammaH} 
\Gamma_H\equiv - \frac{1}{T}\frac{\pd S/\pd B}{\pd S/\pd T}=
-\frac{\pd M/\pd T}{C_H}. \tag{S-1}
\eeq
The divergence of this quantity as $T \to 0$ is a strong indicator of quantum criticality~\cite{Zhu03-S}. Our results for \ybal in Fig. 3 in the main text show a clear divergence of $\Gamma_H/B$ as $T \to 0$ in the non-Fermi liquid regime ($B \lesssim T$), whereas it levels off at low $T$ in the Fermi liquid regime ($T \lesssim B$). 
Note that a small nuclear Schottky contribution to $C$ (see Fig.~1B in the main text) 
may slightly affect the results below $\sim 0.2$ K. 
However, after a subtraction of this nuclear contribution, the $T$ dependence of $\Gamma_H$ becomes more divergent at low fields and temperatures than the results shown 
in Fig. 3 in the main text.
In addition, according to theory, $\Gamma_H$ at a finite temperature is expected to exhibit two types of behavior as a function of field: in the NFL regime ($B < T$), $\Gamma_H\propto |B-B_{\rm c}|$ and in the FL regime ($B > T$), then $\Gamma_H\propto 1/|B-B_{\rm c}|$~\cite{Zhu03-S}. Figure~S2 shows the field dependence of $\Gamma_H$, in which we see two clear regimes, as predicted by theory~\cite{Zhu03-S}. From the NFL regime, we can extract the critical field $\vert B_c \vert < 0.2$~mT, consistent with the estimate of $B_c$ obtained from the scaling behavior of $M$, Eq.1 in the main text.

\subsubsection*{2. Power-law fit to the low-temperature susceptibility}

The previous paper of \ybal, Ref.~S5, has initially reported the $T^{-1/3}$ power-law of the low-temperature magnetic susceptibility. However present measurements, carried out on 
significantly higher quality single crystals, have shown that the exponent $T^{-1/2}$ provides a better fit to the experimental results and is also in agreement with the exponent derived from $T/B$ scaling of magnetization (see next section). The present data more accurately represent the intrinsic behavior in the material, not only because of the higher quality of the single crystals used, but also because greater care was taken in the etching process to fully remove sample surface impurities, and furthermore, the current data were obtained without a drift in the SQUID output and without a sizable background signal. These refinements have been already reported in Ref.~S3.

\subsubsection*{3. $\boldsymbol{(T/B)}$ scaling of thermodynamic quantities}

The scaling property of magnetization $M$ shown in Eq.~1 of the main text,
\beq\label{S.dMdT}
-{\rm d}M_c/{\rm d}T = B^{\alpha-2} \phi(T/B),\tag{S-2}
\eeq 
with $\alpha=3/2$ indicates that all thermodynamic properties can be expressed as a function of the ratio of $T$ and $B$. Integrating both parts over temperature results in the following expression for magnetization:
\beq \label{S.M}
M_{\rm c} \equiv M - M_{\rm 0} = -B^{\alpha-1} \tilde{\phi}(T/B), \tag{S-3} 
\eeq
where $\ud\tilde{\phi}/\ud x = \phi(x)$. Above, $M_{\rm c}$ and $M_{\rm 0}$ is the critical and non-critical components of the magnetization, respectively. For $\alpha=3/2$, this means that $M_{\rm c}/B^{1/2}$ is a universal function of the ratio of ($T/B$) only, as plotted in Fig.~S3.
Since the magnetization $M$ is the derivative of the free energy, $M= -{\rm d}F/{\rm d}B$, integrating both parts of Eq.~\ref{S.M} over field, one obtains the following simple scale-invariant form of the low-temperature free energy of \ybal :
\beq\label{S.free}
F= B^\alpha f\left(\frac{T}{B}\right),\tag{S-4}
\eeq
where $f(x)$ is a scaling function of the ratio $T/B$, which is related to $\phi$ in Eq.~\ref{S.dMdT} as follows: $\phi(x)=(\alpha-1)f'(x) - x f''(x)$. To ensure that the free energy depends only on temperature in the non-Fermi liquid limit $T\gg B$, we require that the function $f(x)\sim x^\alpha$ for $x \gg 1$. In the opposite limit $T\ll B$ the system is a Fermi liquid, and hence one must be able to expand the free energy in powers of $T^2$, which in turn requires that $f(x) \sim {\rm const} + \mathcal{O}(x^2)$ for $x \ll 1$.  In other words, $f(x)$ is required to have the following asymptotic behavior in the two limits:
\beq
f(x)\propto \left\{ \begin{array}{ll}
  x^{\alpha}, & \text{for } x \gg 1 \quad (T\gg B, \text{non-Fermi liquid}) \\
  \text{const} + x^2, & \text{for } x \ll 1 \quad (T\ll B, \text{Fermi liquid})
\end{array} \right.\tag{S-5}
\eeq
The function chosen for the study is $f(x) = -\lambda (A+x^2)^{\alpha/2}$, resulting in the scaling function
\beq \label{S.phi}
\phi(x) = \lambda A\alpha(2-\alpha) x (A+x^2)^{\frac{\alpha}{2}-2} \equiv \Lambda x(A+x^2)^{-n}.\tag{S-6}
\eeq
Figure 2B in the main text shows the fit of the empirical scaling relation Eq.~\ref{S.dMdT} to the above form. The best fit is achieved with $n=1.25 \pm 0.01$ corresponding to $\alpha=1.50 \pm 0.02$, 
in agreement with the experimental data. 
This now allows us to write down the free energy in the simple form
\beq\label{S.newF}
F =  -\frac{1}{(\kB \tilde{T})^{\alpha-1}}\left((g\mu_\text{B}B)^2 + (\kB T)^2 \right)^{\alpha/2},\tag{S-7}
\eeq
where $g$ is the effective $g$-factor of the quantum critical excitations, related to the coefficient $A$ in Eq.~\ref{S.phi} as $A = (g\mu_\text{B}/\kB)^2$, so that the scaling fit results in the effective moment $g\mu_\text{B} \approx 1.94\mu_\text{B}$. The constant energy scale $\tilde{T}$ in the prefactor can be determined from the following identity: $\tilde{T} = (k_\text{B}A\alpha(2-\alpha)/\Lambda)^{\frac{1}{\alpha-1}}$, yielding $\kB\tilde{T}\approx6.56$~eV. While seemingly very large (of the order of the conduction electron bandwidth), this value of $\tilde{T}$ agrees well with the measured value of the specific heat coefficient. Indeed, differentiating Eq.~\ref{S.newF} twice over $T$, one arrives at the specific heat coefficient $C_\text{QC}/T \sim \frac{3R}{4}(T\tilde{T})^{-1/2}$, which at $T=0.1$~K yields a value  77~mJ/mol K$^2$ for the quantum critical component of the specific heat coefficient. 
This value of $C_\text{QC}/T$ is close to an experimental estimate of the QC component of $C/T$ at $T=0.1$~K, $\gamma_\text{QC}\approx 60$~mJ/mol~K$^2$ for  the $\beta$ phase. The estimate is obtained after subtracting $C/T \approx 130$ mJ/mol~K$^2$ for $\alpha$-YbAlB$_4$ at 0.1 K, the sum of the electronic non-critical component and the nuclear Schottky component, both of which are likely similar for these locally isostructural compounds.

 The above simple form of the scaling relation (\ref{S.newF}) is significant in that the quantum critical free energy only depends on the distance from the origin in the $(\kB T,g\mu_\text{B} B)$ coordinates, similar to e.g. the case of the Tomonaga--Luttinger liquid, where the free energy takes a similar form $F_\text{TL}\propto (\kB T)^2 + (\tilde{g}\mu_\text{B}B)^2$ with a model-dependent $g$-factor $\tilde{g}$ \cite{sachdev-S}.

The above scaling relations, Eqs.~\ref{S.dMdT}, \ref{S.free} can be easily modified to a more general case of non-zero critical field by setting $B\to B-B_c$ in the above expression. This allowed us to place an upper bound on the value of the critical field, $\vert B_c \vert <0.2$ mT, deduced from Pearson's fitting quality shown in the inset of Fig. 2B, as discussed in the main text.

The field dependence of the specific heat coefficient $\gamma(T) = C(T)/T = - {\pd^2}F/{\pd}T^2$ can be deduced from Eq.~\ref{S.free} to be 
\beq
\gamma(T,B) =  -B^{\alpha -2}f''(T/B).\tag{S-8}
\eeq
The limiting $T \to 0$ behavior of this quantity is proportional to the effective mass of the quasi-particles: $m^*(B) \propto \gamma(T=0, B)$, giving 
\beq
m^*(B) \propto B^{\alpha -2}f''(0),\tag{S-9}
\eeq
which yields $m^* \propto B^{-1/2}$ for the empirically determined value of $\alpha = 1.5$.

$T/B $ scaling in \ybal is observed over an
unprecedentedly wide range of more than three decades in the non-Fermi
liquid regime $T/B>1$. This may be contrasted with other known
examples of $T/|B-B_c|$ scaling in quantum critical heavy fermion
materials such as \CeCuAu ($B_c=0$) and \yrs ($B_c\neq0$), where in
the non-Fermi liquid regime, the scaling  was observed over roughly
one decade of this ratio~\cite{schroeder-S,Custers03-S}. Known
theoretical examples of quantum models displaying $T/B$
scaling include~\cite{sachdev-S}: the Tomonaga--Luttinger liquid in one-dimensional
metals, the critical (2+1)-dimensional $\mathcal{O}(N)$
sigma model and a drained Fermi liquid with a chemical potential $\mu=0$.

Currently, it is not possible to extract the pure electronic contribution to the specific heat at zero field because the nuclear contribution is unknown. However, we may analyze the change in the specific heat under application of a low field using 
the Maxwell relation $({\pd}S/{\pd}B)_T = ({\pd}M/{\pd}T)_B$ and the scaling equation (\ref{S.dMdT}). 
Based on the Maxwell relation, the specific heat $C(B)$ in magnetic field $B$ can be expressed as follows, 
\beq\label{S.Maxwell}
\frac{C(B)}{T}=\int_0^B \frac{{\pd^2}M}{{\pd}T^2}dB+\frac{C(0)}{T}.\tag{S-10}
\eeq
Here, $C(0)$ is the zero field specific heat. Figure S4 shows that the calculated results using 

Eq.~\ref{S.dMdT} and Eq.~\ref{S.Maxwell} are 
in the full agreement with the experimental results for $B =$ 0.07 and 0.5 T. 
This indicates that the free energy given in Eq.~2 in the main text and Eq.~\ref{S.free} is consistent with the temperature 
and field dependence of the specific heat in the quantum critical regime.

\clearpage

\begin{figure}[tb]
\begin{center}
\includegraphics[width = 9 cm,clip]{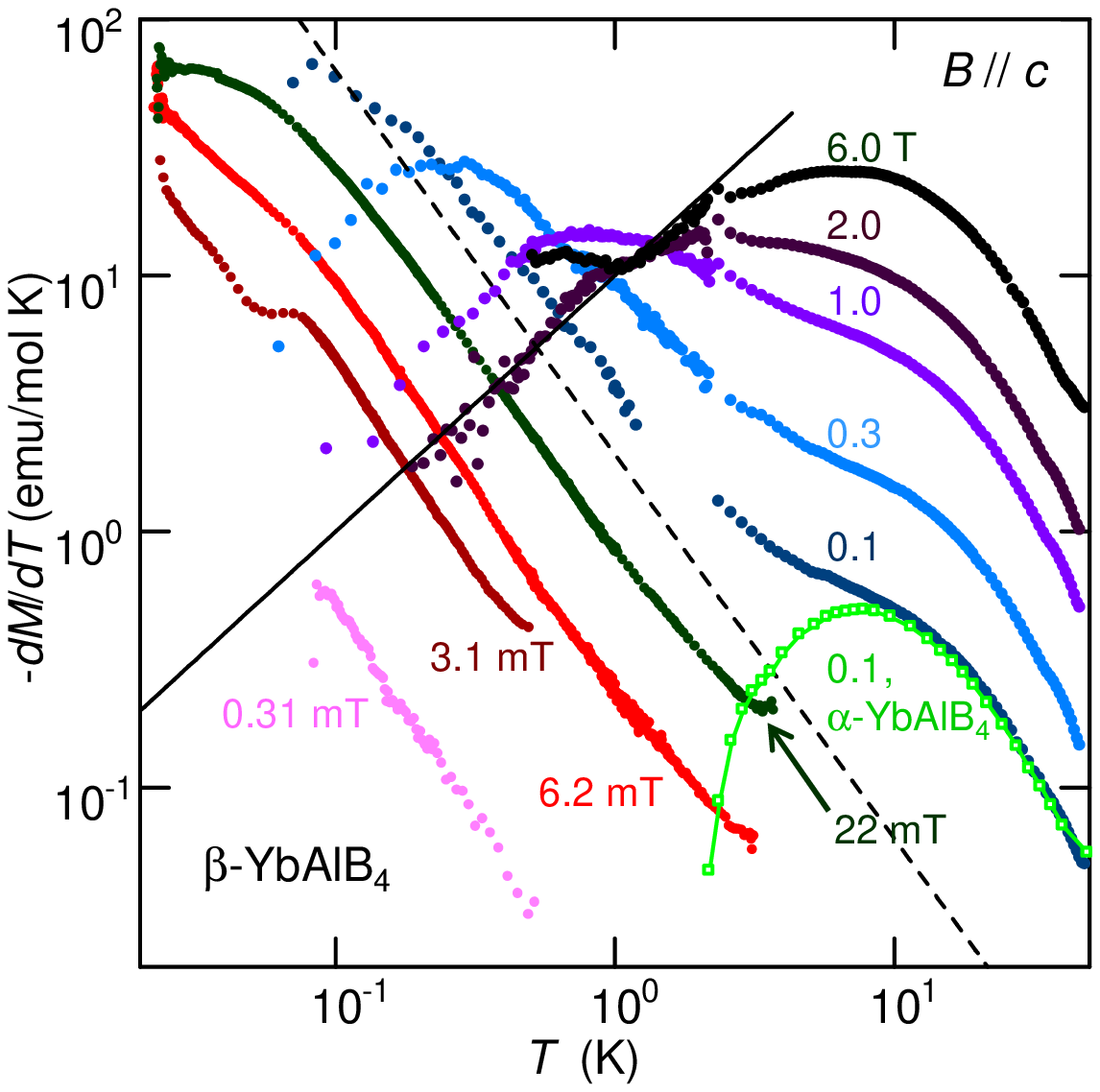}
\end{center}
\caption{$-dM/dT$ versus $T$ for selected fields on a logarithmic scale for both $\beta$- (solid circles) and $\alpha$-YbAlB$_4$ (open squares). The sudden downturn below 0.1 K in the low field data corresponds to the onset of superconductivity.}
\label{fS1}
\end{figure}

\begin{figure}[tb]
\begin{center}
\includegraphics[width = 11 cm,clip]{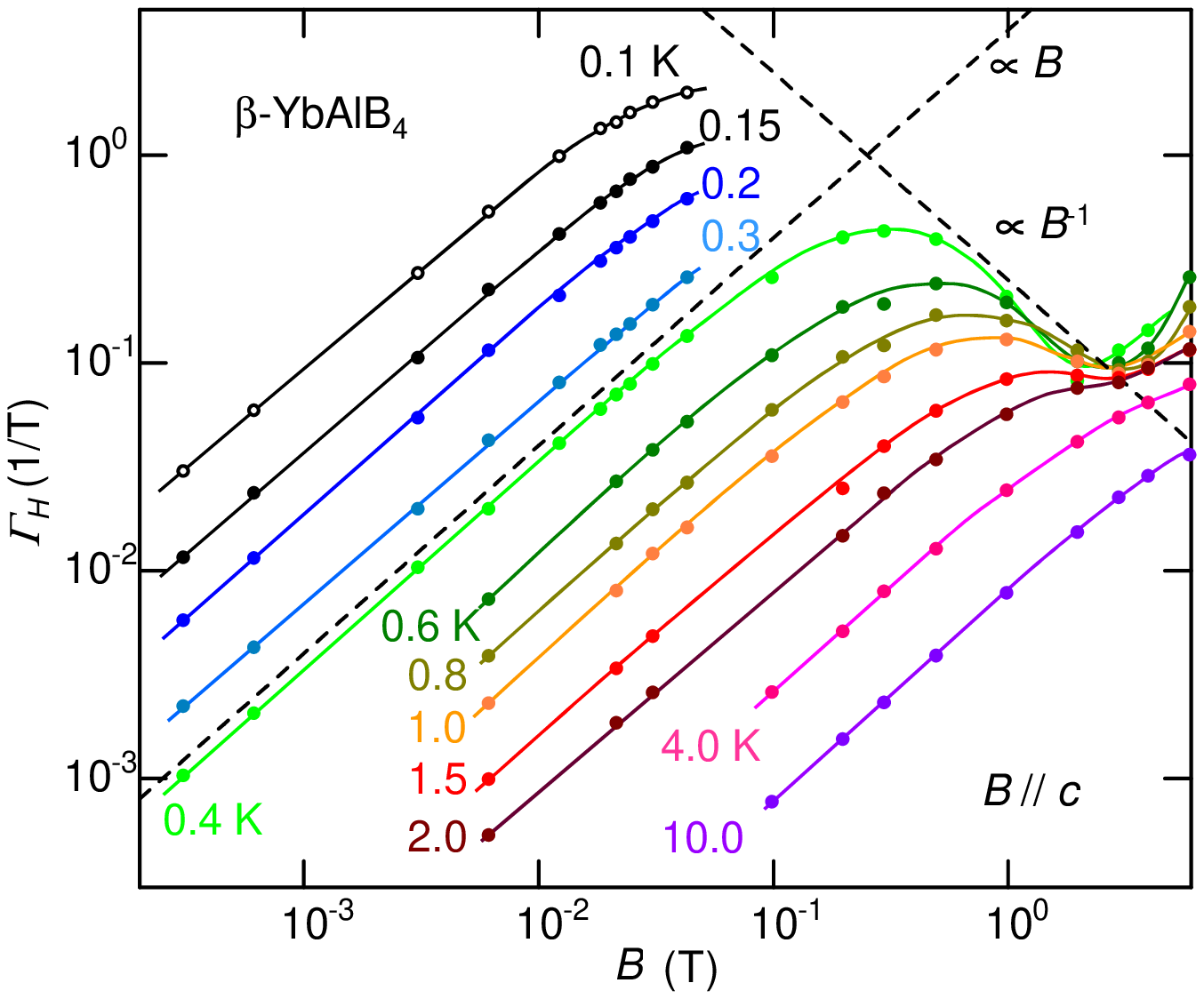}
\end{center}
\caption{Magnetic field dependence of the magnetocaloric effect for \ybal, 
$\Gamma_H\equiv -(\pd M/\pd T)/C$, obtained from $ -dM/dT$ and the total specific heat $C$. 
In the NFL regime at $T > B$, $\Gamma _H$ increases linearly with $B$, while in the FL regime at $T < B$, it exhibits $B^{-1}$ dependence, consistent with theory \cite{Zhu03-S}. The upturn observed at $\sim$ 2 T in the temperature range of 0.4 K $< T <$ 1.5 K indicates the boundary between the quantum critical and non-quantum critical regions.}
\label{fS2}
\end{figure}

\begin{figure}[tb]
\begin{center}
\includegraphics[width = 11 cm,clip]{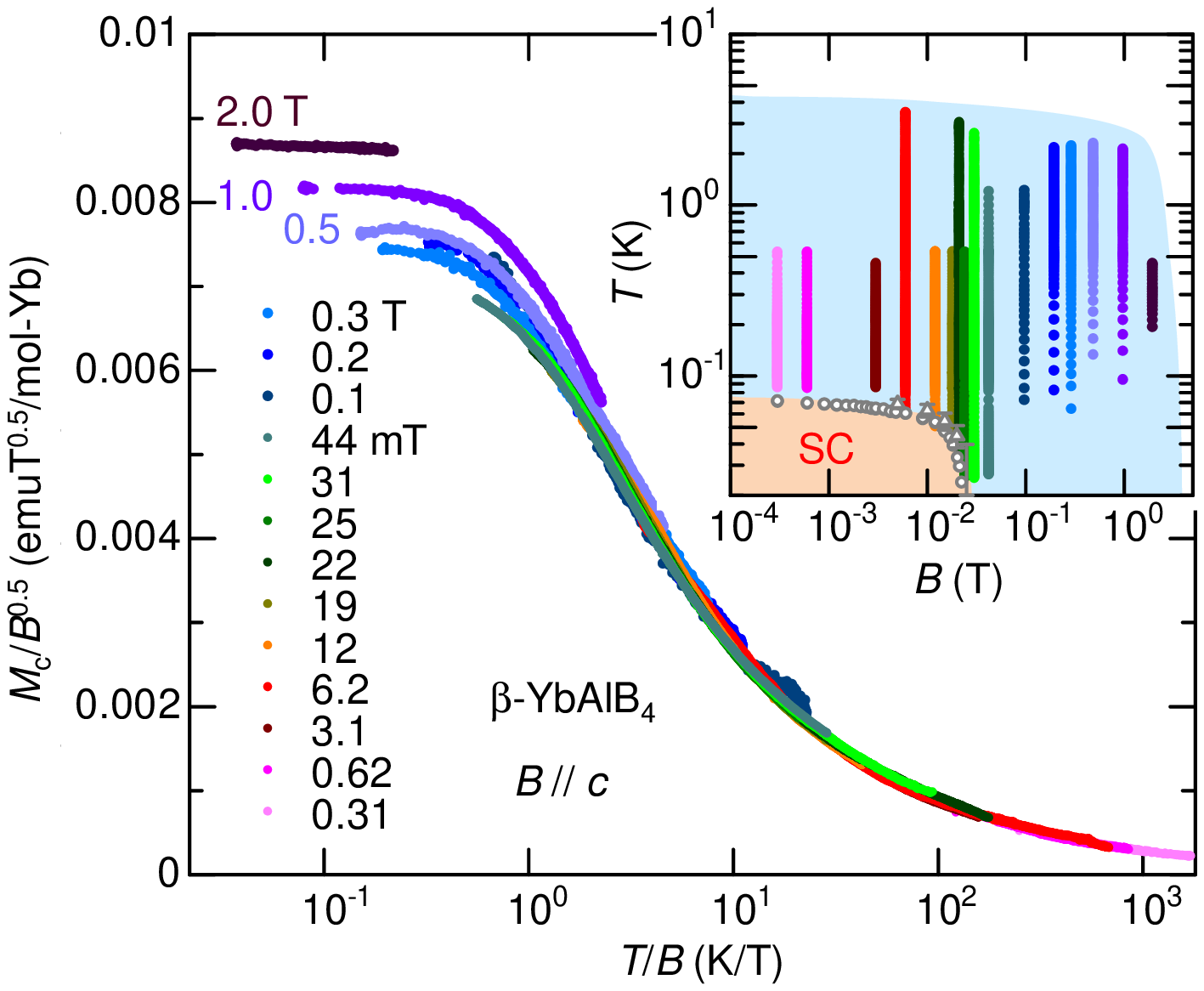}
\end{center}
\caption{Scaling plot of $M_c/\sqrt{B}$ vs. ratio $T/B$ for \ybal, 
which according to Eq.~(\ref{S.M}) should be a universal function of this ratio. Here, $M_c$ is the critical contribution to the magnetization: \mbox{$M_c = M - M_0$}, where we have subtracted $M_0 = \chi_0 B$ using the constant non-critical susceptibility $\chi_0 = 0.017$ emu/mol. This latter constant term, whose nature is immaterial to the scaling analysis, is close to the zero-$T$ susceptibility of the non-critical $\alpha$-YbAlB$_4$. It may originate from constant Van Vleck contribution to susceptibility and/or from Pauli susceptibility of the non-critical parts of the Fermi surfaces. Such a subtraction is well justified in the literature and was used for instance to establish the $T/B$ scaling in CeCu$_{6-x}$Au$_x$ (see Fig. 4b in Ref.~\cite{schroeder-S}). Under fields $B > 0.5$ T, $M_{\rm c}$ does not follow a single scaling curve, because of a small field-nonlinear contribution to $M_0$, which is neglected 
in the analysis. Inset shows the quantum critical $B$-$T$ range where the scaling applies (solid
circles in the blue shaded region) and the superconducting phase boundary (open circles\cite{matsumoto} and open triangles\cite{KugaPRL-S}) of \ybal.}
\label{fS3}
\end{figure}

\begin{figure}[tb]
\begin{center}
\includegraphics[width = 9 cm,clip]{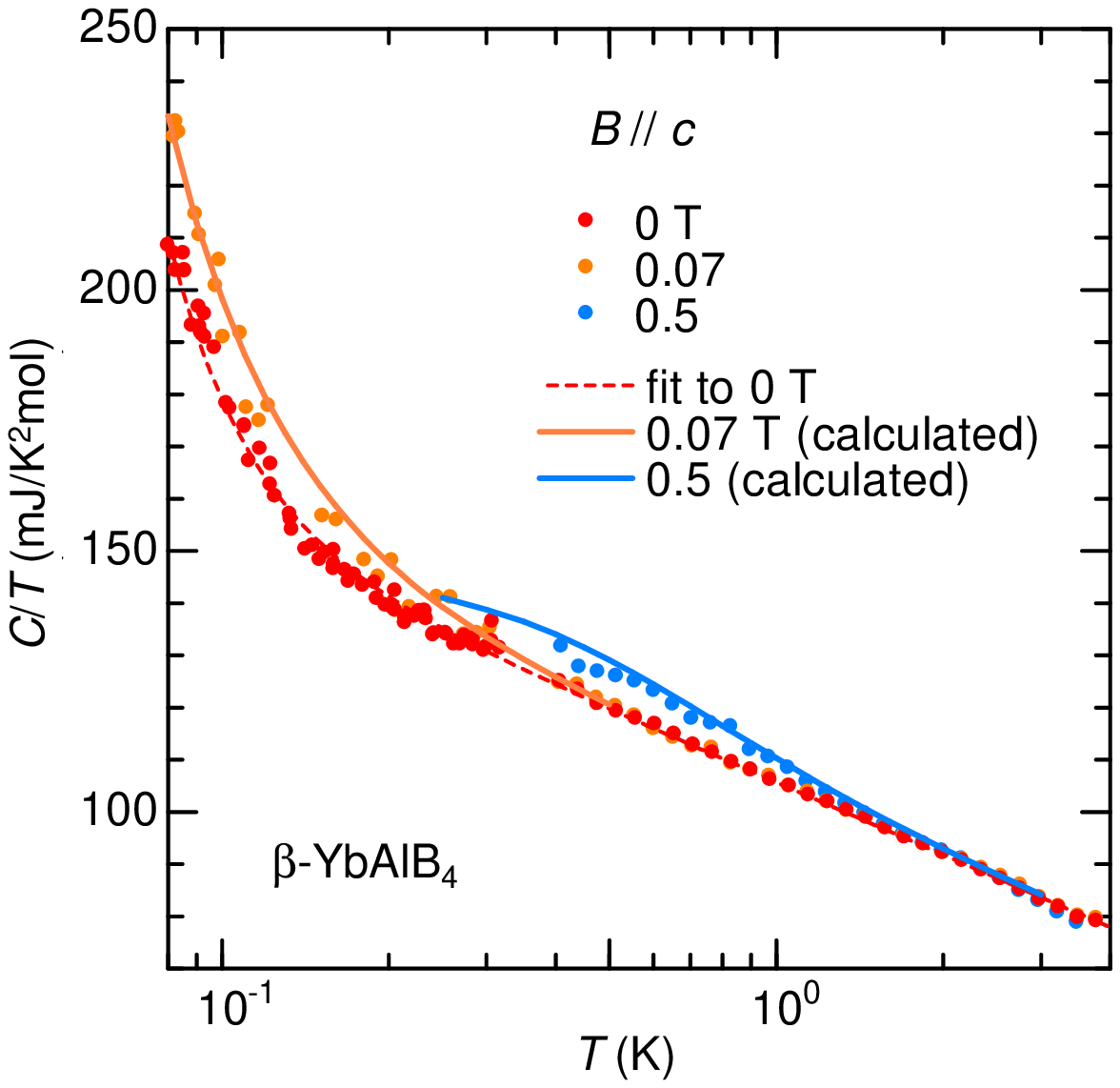}
\end{center}
\caption{Specific heat divided by temperature $C/T$ in the low fields below 0.5 T along the $c$ axis (filled symbols) 
and those estimated by using the scaling relation, Eq. (1) in the main text (or Eq.(\ref{S.dMdT})) as well as the Maxwell relation, Eq.~(\ref{S.Maxwell}) (solid lines, see text). 
The experimental results and the calculation agree well with each other in the low field below 0.5 T in the quantum critical regime.
The broken line is a fit to the zero filed data $C(0)/T$, which was used in Eq.~(\ref{S.Maxwell}).  }
\label{fS4}
\end{figure}

\end{document}